\documentstyle[12pt]{article}
\input psfig

\def\pasp{{Pub. Astron. Soc. Pac.}}
\def\apj{{Astrophys. J.}}
\def\apjl{{Astrophys. J. Lett.}}
\def\prl{{Phys. Rev. Lett.}}
\def\prd{{Phys. Rev. D}}
\def\mnras{{Mon. Not. R. astron. Soc.}}
\def\aj{{Astron. J.}}
\def\araa{{Ann. Rev. Astron. Astrophys.}}

\def\cmm2{{\,\rm cm^{-2}}}
\def\cm2{{\,{\rm cm}^2}}
\def\cmm3{{\,{\rm cm}^{-3}}}
\def\gcmm3{{\,{\rm g\,cm^{-3}}}}

\def\fun#1#2{\lower3.6pt\vbox{\baselineskip0pt\lineskip.9pt
  \ialign{$\mathsurround=0pt#1\hfil##\hfil$\crcr#2\crcr\sim\crcr}}}

\begin{document}
\pagestyle{empty}
\begin{center}
\bigskip


\vspace{1in}
{\Large \bf Dark Matter and Energy in the Universe
\footnote{To appear in {\em Physica Scripta},
Proceedings of the Nobel Symposium, Particle Physics
and the Universe (Enkoping, Sweden, August 20-25, 1998)}}
\bigskip

\vspace{.2in}
Michael S. Turner\\

\vspace{.2in}
{\it Departments of Astronomy \& Astrophysics and of Physics\\
Enrico Fermi Institute, The University of Chicago, Chicago, IL~~60637-1433}\\

\vspace{0.1in}
{\it NASA/Fermilab Astrophysics Center\\
Fermi National Accelerator Laboratory, Batavia, IL~~60510-0500}\\

\end{center}

\vspace{.3in}
\centerline{\bf ABSTRACT}

\bigskip

For the first time, we have a plausible and complete accounting
of matter and energy in the Universe.  Expressed a fraction
of the critical density it goes like this:  neutrinos, between 0.3\%
and 15\%; stars, between 0.3\% and 0.6\%; baryons (total), $5\% \pm 0.5\%$;
matter (total), $40\% \pm 10\%$; smooth, dark energy, $80\% \pm 20\%$;
totaling to the critical density (within the errors).
This accounting is consistent with the inflationary prediction
of a flat Universe and defines three dark-matter problems:
Where are the dark baryons?  What is the nonbaryonic dark matter?
What is the nature of the dark energy?  The leading candidate
for the (optically) dark baryons
is diffuse hot gas; the leading candidates for the nonbaryonic
dark matter are slowly moving elementary particles left over from
the earliest moments (cold dark matter), such as axions or
neutralinos; the leading candidates for the dark energy involve
fundamental physics and include a
cosmological constant (vacuum energy), a rolling scalar field
(quintessence), and a network of light, frustrated topological defects.

\newpage
\pagestyle{plain}
\setcounter{page}{1}

\section{Introduction}

The quantity and composition of matter and energy in the
Universe is of fundamental importance in cosmology.  The
fraction of the critical energy density contributed by
all forms of matter and energy,
\begin{equation}
\Omega_0 \equiv {\rho_{\rm tot} \over \rho_{\rm crit}}
= \sum_i \Omega_i \,,
\end{equation}
determines the geometry of the Universe:
\begin{equation}
R_{\rm curv}^2 = {H_0^{-2} \over \Omega_0 - 1}\,.
\end{equation}
Here, subscript `0' denotes the value at
the present epoch, $\rho_{\rm crit} = 3H_0^2/8\pi G \simeq 1.88h^2\times
10^{-29}\,{\rm g\ cm^{-3}}$, $\Omega_i$ is the fraction of
critical density contributed by component $i$ (e.g.,
baryons, photons, stars, etc) and $H_0 = 100h\,{\rm km\,s^{-1}\,
Mpc^{-1}}$.  The sign of $R_{\rm curv}^2$ specifies the spatial
geometry:  positive for a 3-sphere, negative for
a 3-saddle and 0 for the flat space.

The present rate of deceleration of the expansion depends upon $\Omega_0$
as well as the composition of matter and energy,
\begin{equation}
q_0 \equiv {(\ddot R /R)_0 \over H_0^2}
        = {1 \over 2}\Omega_0 + {3\over 2} \sum_i \Omega_i w_i \,.
\end{equation}
The pressure of component $i$, $p_i \equiv w_i \rho_i$;
e.g., for baryons $w_i = 0$, for radiation $w_i = 1/3$,
and for vacuum energy $w_i = -1$.

The fate of the Universe -- expansion forever or recollapse --
is not directly determined by $\Omega_0$ and/or $q_0$.
It depends upon precise knowledge of the composition of {\em all}
components of matter and energy, for all times in the future.
Recollapse occurs only if there is a future turning point, that is
an epoch when the expansion rate,
\begin{equation}
H^2 = {8\pi G \over 3} \sum_i \rho_i - {1\over R_{\rm curv}^2}\,,
\end{equation}
vanishes and
\begin{equation}
{\ddot R \over R} = -{4\pi G\over 3} \sum_i\,\rho_i[1+w_i]
\end{equation}
is less than zero.  In a universe comprised
of matter alone, a positively curved universe ($\Omega_0 > 1$)
eventually recollapses and a negatively
curved universe ($\Omega_0 < 1$) expands forever.  However,
exotic components complicate matters:  a positively curved
universe with positive vacuum energy can expand forever, and
a negatively curved universe with negative vacuum energy can recollapse.

The quantity and composition of matter and energy in the Universe
is also crucial for understanding the past.  It
determines the relationship between age of the Universe
and redshift, when the Universe ended
its early radiation dominated era, and the growth of small inhomogeneities
in the matter.  Ultimately, the formation and evolution of
large-scale structure and even individual galaxies depends upon
the composition of the dark matter and energy.

Measuring the quantity and composition of matter and energy in the Universe
is a challenging task.  Not just because the scale of inhomogeneity
is so large, around $10\,$Mpc; but also, because there may be components
that remain exactly or relatively smooth (e.g., vacuum energy or
relativistic particles) and only reveal their presence by their
influence on the evolution of the Universe itself.

\section{A Complete Inventory of Matter and Energy}

\subsection{Preliminaries}
\subsubsection{Radiation}

Because the cosmic background radiation (CBR)
is known to be black-body radiation to very high precision
(better than $0.005\%$) and its temperature is known to four
significant figures, $T_0 = 2.7277\pm 0.002\,$K, its
contribution is very precisely known, $\Omega_\gamma
h^2 = 2.480\times 10^{-5}$.
If neutrinos are massless or very light, $m_\nu \ll 10^{-4}\,$eV,
their energy density is equally well known because it is directly
related to that of the photons, $\Omega_\nu = {7\over 8}(4/11)^{4/3}
\Omega_\gamma$ (per species) (there is a small 1\% positive
correction to this number; see Dodelson \& Turner, 1992).

It is possible that additional relativistic species contribute
significantly to the energy density, though big-bang nucleosynthesis (BBN)
severely constrains the amount (the equivalent of less than
0.4 of a neutrino species; see e.g., Burles et al, 1999)
unless they were produced by the decay of a massive particle
after the epoch of BBN.

In any case, we can be confident that the radiation component of
the energy density today is small.  
The matter contribution (denoted by $\Omega_M$), consisting of particles
that have negligible pressure, is the easiest to determine because
matter clumps and its gravitational effects are thereby enhanced
(e.g., in rich clusters the matter density typically exceeds
the mean density by a factor of 1000 or more).
With this in mind, I will decompose the present matter/energy
density into two components, matter and vacuum energy,
\begin{equation}
\Omega_0 = \Omega_M + \Omega_\Lambda\,,
\end{equation}
ignoring the contribution of the CBR
and ultrarelativistic neutrinos.
I will use vacuum energy
as a stand in for a smooth component (more later).  Vacuum energy
and a cosmological constant are indistinguishable:  a cosmological
constant corresponds to a uniform energy density of magnitude
$\rho_{\rm vac} = \Lambda /8\pi G$.

\subsection{$\Omega_0 = 1\pm 0.2$}

There is a growing consensus that the anisotropy of the CBR
offers the best means of determining the curvature of the Universe
and thereby $\Omega_0$, cf., Eq. (2).
This is because the method is intrinsically geometric --
a standard ruler on the last-scattering surface --
and involves straightforward physics at
a simpler time (see e.g., Kamionkowski et al, 1994).

\begin{figure}
\centerline{\psfig{figure=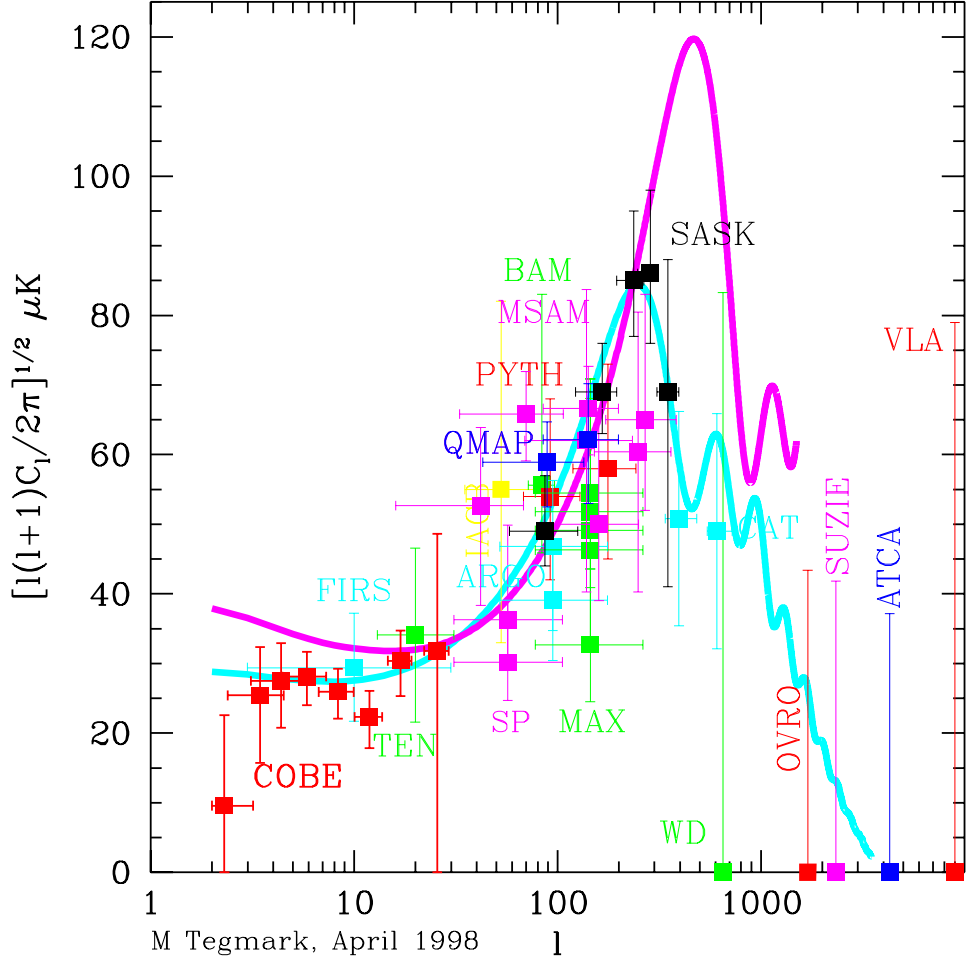,width=5in}}
\caption{Summary of all CBR anisotropy measurements, where
the temperature variations across the sky have been expanded
in spherical harmonics, $\delta T(\theta , \phi ) = \sum_i a_{lm}Y_{lm}
(\theta ,\phi )$
and $C_l \equiv \langle |a_{lm}|^2\rangle$.  In plain
language, this plot shows the size of the temperature variation
between two points on the sky separated by angle $\theta$
(ordinate) vs. multipole number $l=200^\circ / \theta$
($l=2$ corresponds to $100^\circ$, $l=200$ corresponds to $\theta = 1^\circ$,
and so on).  The curves illustrate the predictions of CDM models
with $\Omega_0 = 1$ (curve with lower peak) and $\Omega_0 =0.3$ (darker
curve).  Note:  the preference of the data for a flat Universe and
the evidence for the first of a series of ``acoustic peaks.''
The presence of these acoustic peaks is a key signature
of the density perturbations of quantum origin predicted by inflation
(Figure courtesy of M. Tegmark).}
\label{fig:cbr_today}
\end{figure}

At last scattering baryonic matter (ions and electrons) was still tightly
coupled to photons; as the baryons fell into the dark-matter
potential wells the pressure of photons acted as a restoring
force, and gravity-driven acoustic oscillations resulted.  These
oscillations can be decomposed into their Fourier modes;
Fourier modes with $k\sim l H_0/2$ determine the multipole
amplitudes $a_{lm}$ of CBR anisotropy.  Last scattering
occurs over a short time, making the CBR is a snapshot of
the Universe at $t_{\rm ls} \sim 300,000\,$yrs.
Each mode is ``seen'' in a well defined phase of its
oscillation.  (For the density perturbations predicted by inflation,
all modes the have same initial phase because
all are growing-mode perturbations.)
Modes caught at maximum compression or rarefaction lead
to the largest temperature anisotropy; this results in a series of
acoustic peaks beginning at $l\sim 200$ (see Fig.~\ref{fig:cbr_today}).
The  wavelength of the lowest
frequency acoustic mode that has reached maximum compression,
$\lambda_{\rm max} \sim v_s t_{\rm ls}$, is the standard
ruler on the last-scattering surface.  Both $\lambda_{\rm
max}$ and the distance to the last-scattering surface depend
upon $\Omega_0$, and the position of the first peak $l\simeq
200/\sqrt{\Omega_0}$.  This relationship is insensitive
to the composition of matter and energy in the Universe
(see Spergel, 1999).

\begin{figure}
\centerline{\psfig{figure=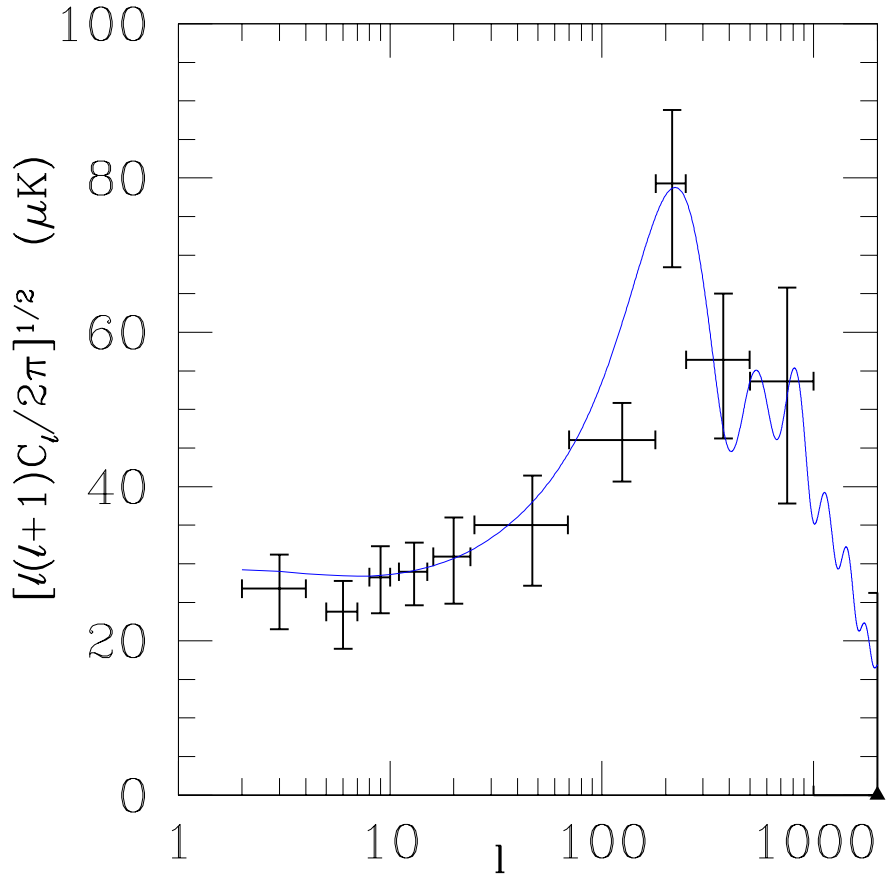,width=5in}}
\caption{The same data as in Fig.~2, but
averaged and binned to reduce error bars and
visual confusion.  The theoretical
curve is for the $\Lambda$CDM model with $H_0=65\,{\rm km\,
s^{-1}\,Mpc^{-1}}$ and $\Omega_M =0.4$; note the
goodness of fit (Figure courtesy of L. Knox).
}
\label{fig:cbr_knox}
\end{figure}

CBR anisotropy measurements, shown in Figs.~\ref{fig:cbr_today} and
\ref{fig:cbr_knox}, now cover
three orders of magnitude in multipole number and involve
more than twenty experiments.  COBE is
the most precise and covers multipoles $l=2-20$;
the other measurements come from balloon-borne, Antarctica-based
and ground-based experiments using both low-frequency
($f<100\,$GHz) HEMT receivers and high-frequency ($f>100\,$GHz)
bolometers.  Taken together, all the measurements are beginning to
define the position of the first acoustic peak, at a value that is
consistent with a flat Universe.  Various analyses of the
extant data have been carried out, indicating $\Omega_0 \sim 1\pm 0.2$
(see e.g., Lineweaver, 1998).
It is certainly too early to draw definite conclusions or put
too much weigh in the error estimate.  However, a strong
case is developing for a flat Universe and more data is on
the way (Python V, Viper, MAT, Maxima, Boomerang, DASI, and others).
Ultimately, the issue will be settled by NASA's
MAP (launch late 2000) and ESA's Planck (launch 2007) satellites
which will map the entire CBR sky with 30 times the resolution
of COBE (around $0.1^\circ$) (see Wilkinson, 1999).

\subsection{Matter}

\subsubsection{Baryons}

For more than twenty years big-bang nucleosynthesis has provided
a key test of the hot big-bang cosmology as well as
the most precise determination of the baryon density.
Careful comparison of the primeval abundances of D, $^3$He, $^4$He
and $^7$Li with their big-bang predictions defined a
concordance interval, $\Omega_Bh^2 = 0.007 - 0.024$ (see e.g.,
Copi et al, 1995; for another view, see Hata et al, 1995).

Of the four light elements produced in the big bang,
deuterium is the most powerful ``baryometer'' -- its primeval abundance
depends strongly on the baryon density ($\propto 1/\rho_B^{1.7}$) --
and the evolution of its abundance since the big bang is simple --
astrophysical processes only destroy deuterium.  Until recently
deuterium could not be exploited as a baryometer because its
abundance was only known locally, where roughly half of the material
has been through stars with a similar amount of
the primordial deuterium destroyed.  In 1998, the situation changed
dramatically, launching BBN into the precision era of cosmology.

Over the past four years there have been claims of upper limits,
lower limits, and determinations
of the primeval deuterium abundance, ranging from (D/H)\,$=10^{-5}$
to (D/H)\,$=3\times 10^{-4}$.  In short, the situation was
confusing.  In 1998 Burles and Tytler clarified matters and established
a strong case for (D/H)$_P = (3.4\pm 0.3)\times 10^{-5}$
(Tytler, 1999).  That case is based upon the deuterium abundance measured
in four high-redshift hydrogen clouds seen in absorption against distant QSOs,
and the remeasurement and reanalysis of other putative deuterium systems.
In this enterprise, the Keck I 10-meter telescope and its HiRes Echelle
Spectrograph have played the crucial role.

\begin{figure}
\centerline{\psfig{figure=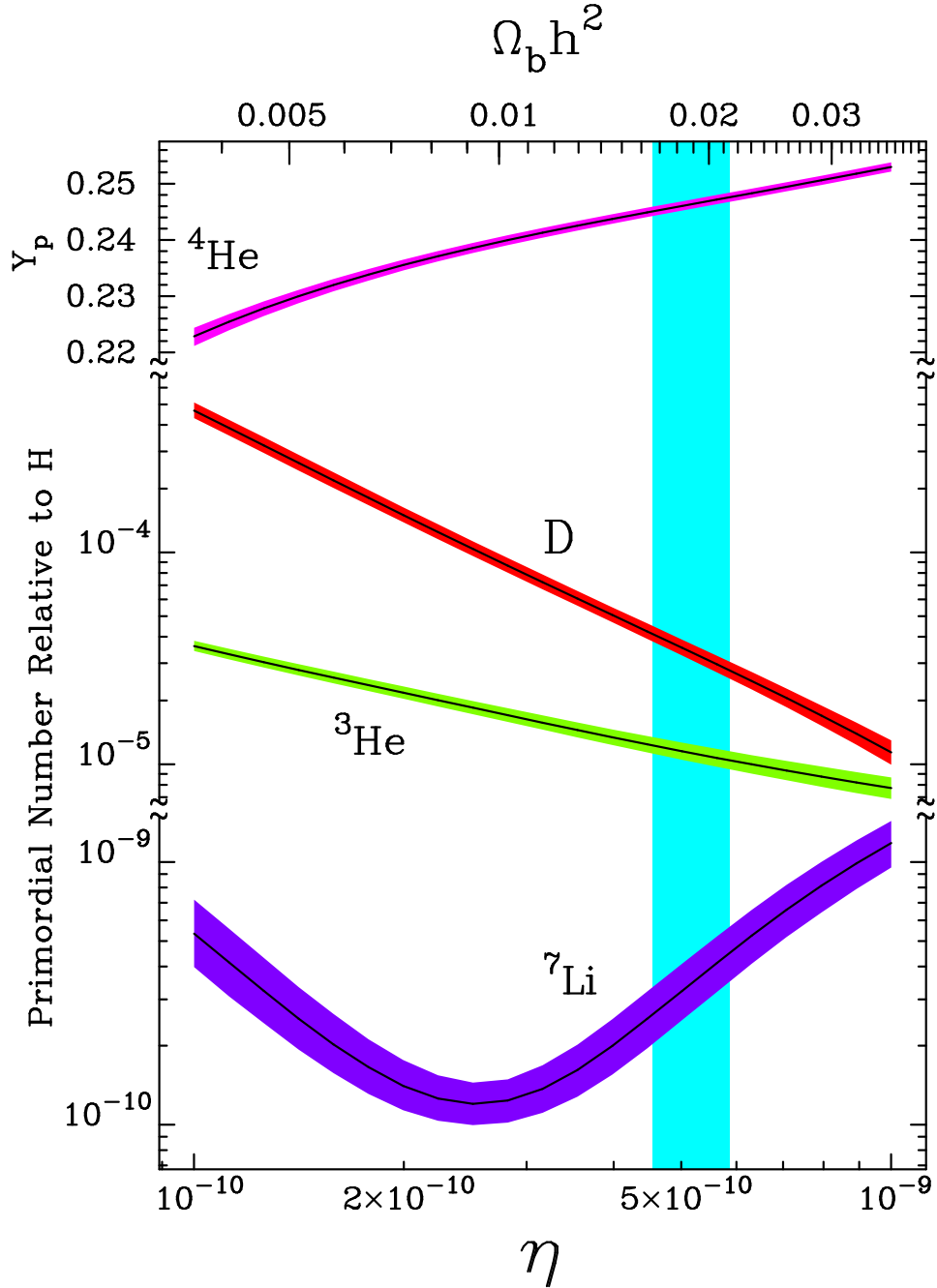,width=5in}}
\caption{Predicted abundances of $^4$He (mass fraction),
D, $^3$He, and $^7$Li (number relative to hydrogen) as a function
of the baryon density; widths of the curves indicate
``$2\sigma$'' theoretical uncertainty.  The dark band highlights
the determination of the baryon density based upon
the recent measurement of the primordial abundance of
deuterium (Burles \& Tytler, 1998a,b), $\Omega_Bh^2 = 0.019 \pm
0.0024$ (95\% cl); the baryon density is related to the baryon-to-photon
ratio by $\rho_B = 6.88\eta \times 10^{-22}\gcmm3$ (from Burles et al, 1999).
}
\label{fig:bbn}
\end{figure}

The Burles -- Tytler measurement turns the previous factor of three
concordance range for the baryon density into
a 10\% determination of the baryon density, $\rho_B =
(3.8\pm 0.4)\times 10^{-31}\,{\rm g\,cm^{-3}}$ or $\Omega_Bh^2 =
0.02 \pm 0.002$ (see Fig.~\ref{fig:bbn}), with about half
the error in $\rho_B$ coming from the theoretical error in
predicting the BBN yield of deuterium.  [A very recent analysis
has reduced the theoretical error significantly, and improved
the accuracy of the determination of the baryon density,
$\Omega_Bh^2 = 0.019\pm 0.0012$ (Burles et al, 1999).]

This precise determination of
the baryon density, based upon the early Universe physics
of BBN, is consistent with two other measures of the baryon
density, based upon entirely different physics.
By comparing measurements of the opacity of the Lyman-$\alpha$
forest toward high-redshift quasars with high-resolution, hydrodynamical
simulations of structure formation, several groups (Meiksin \& Madau, 1993;
Rauch et al, 1997; Weinberg et al, 1997)
have inferred a lower limit to the baryon density,
$\Omega_Bh^2 > 0.015$ (it is a lower limit because it depends
upon the baryon density squared divided by the
intensity of the ionizing radiation field).
The second test involves the height of the first acoustic peak:
it rises with the baryon density (the higher the baryon density,
the stronger the gravitational force driving the acoustic
oscillations).  Current CBR measurements are consistent with
the Burles -- Tytler baryon density; the MAP and
Planck satellites should ultimately provide a 5\% or better
determination of the baryon density, based upon the
physics of gravity-driven acoustic oscillations when the Universe
was 300,000\,yrs old (see e.g., Spergel, 1999).
This will be an important cross check of the BBN determination.

\subsubsection{Weighing the dark matter:  $\Omega_M = 0.4\pm 0.1$}

Since the pioneering work of Fritz Zwicky and Vera Rubin, it has been
known that there is far too little material in the form of stars
(and related material)
to hold galaxies and clusters together, and thus, that
most of the matter in the Universe is dark (see e.g. Trimble, 1987).
Weighing the dark matter has been the challenge.
At present, I believe that clusters provide the most reliable
means of estimating the total matter density.

Rich clusters are relatively rare objects -- only
about 1 in 10 galaxies is found in a rich cluster --
which formed from density perturbations of (comoving) size
around 10\,Mpc.  However, because they gather together material
from such a large region of space, they can provide
a ``fair sample'' of matter in the Universe.  Using clusters as such,
the precise BBN baryon density can be used
to infer the total matter density (White et al, 1993).
(Baryons and dark matter need not be
well mixed for this method to work
provided that the baryonic and total mass are determined
over a large enough portion of the cluster.)

Most of the baryons in clusters reside in the
hot, x-ray emitting intracluster gas and not in the galaxies
themselves, and so the problem essentially reduces to determining
the gas-to-total mass ratio.
The gas mass can be determined by two methods:
1) measuring the x-ray flux from the intracluster
gas and 2) mapping the Sunyaev - Zel'dovich
CBR distortion caused by CBR photons scattering off hot electrons in the
intracluster gas.  The total cluster mass can be determined
three independent ways:  1)  using the motions of clusters galaxies
and the virial theorem; 2) assuming that the gas is in hydrostatic
equilibrium and using it to infer the underlying mass distribution; and
3) mapping the cluster mass directly by gravitational lensing
(Tyson, 1999).  Within their
uncertainties, and where comparisons can be made, the three methods
for determining the total mass agree (see e.g., Tyson, 1999);
likewise, the two methods for determining the gas mass are consistent.

Mohr et al (1998) have compiled the gas to total mass ratios determined from
x-ray measurements for a sample of 45 clusters; they find
$f_{\rm gas} = (0.075\pm 0.002)h^{-3/2}$.  Carlstrom (1999), using his
S-Z gas measurements and x-ray measurements for the total
mass for 27 clusters, finds $f_{\rm gas} =(0.06\pm 0.006)h^{-1}$.
(The agreement of these two numbers means that clumping of
the gas, which could lead to an overestimate of the gas fraction
based upon the x-ray flux, is not a problem.)
Invoking the ``fair-sample assumption,'' the mean
matter density in the Universe can be inferred:
\begin{eqnarray}
\Omega_M = \Omega_B/f_{\rm gas} & = & (0.3\pm 0.05)h^{-1/2}\  ({\rm X ray})\nonumber\\
                   & = &  (0.25\pm 0.04)h^{-1}\ ({\rm S-Z}) \nonumber \\
                   & = & 0.4\pm 0.1\ ({\rm my\ summary})\,.
\end{eqnarray}
I believe this to be the most reliable and precise
determination of the matter density.  It involves few assumptions,
and most of them have now been tested (clumping, hydrostatic equilibrium,
variation of gas fraction with cluster mass).

\subsubsection{Supporting evidence for $\Omega_M=0.4\pm 0.1$}

This result is consistent with
a variety of other methods that involve different physics.
For example, based upon the evolution
of the abundance of rich clusters with redshift, Henry (1998)
finds $\Omega_M
= 0.45 \pm 0.1$ (also see, Bahcall \& Fan, 1998 and N. Bahcall, 1999).
Dekel and Rees (1994) place a low limit $\Omega_M > 0.3$ (95\% cl)
derived from the outflow of material from voids (a void
effectively acts as a negative mass proportional to the
mean matter density).

The analysis of the peculiar velocities of
galaxies provides an important probe of the mass density averaged
over very large scales (of order several hundred Mpc).  By comparing
measured peculiar velocities with those predicted from the
distribution of matter revealed by redshift surveys such as
the IRAS survey of infrared galaxies,
one can infer the quantity $\beta = \Omega_M^{0.6}/b_I$
where $b_I$ is the linear bias factor that relates the inhomogeneity
in the distribution of IRAS galaxies to that in the distribution of matter
(in general, the bias factor is expected to be in the
range 0.7 to 1.5; IRAS galaxies are expected to be less
biased, $b_I\approx 1$.).  Recent work by
Willick \& Strauss (1998) finds $\beta = 0.5\pm 0.05$, while Sigad et al
(1998) find $\beta = 0.9\pm 0.1$.  The apparent inconsistency of
these two results and the ambiguity introduced by bias
preclude a definitive determination
of $\Omega_M$ by this method.  However, Dekel (1994) quotes a 95\%
confidence lower bound, $\Omega_M >0.3$, and the work of Willick
\& Strauss seems to strongly indicate that $\Omega_M$ is much less than 1;
both are consistent with $\Omega_M\sim 1$.

Finally, there is strong, but circumstantial, evidence from structure formation
that $\Omega_M$ is around 0.4 and significantly greater than $\Omega_B$.
It is based upon two different lines of reasoning.  First,
there is no viable model for structure formation
without a significant amount of nonbaryonic dark matter.
The reason is simple:  in a baryons-only
model, density perturbations grow only from the time of decoupling, $z\sim
1000$, until the Universe becomes curvature dominated, $z\sim \Omega_B^{-1}
\sim 20$; this is not enough growth to produce all the structure
seen today with the size of density perturbations inferred from
CBR anisotropy.  With nonbaryonic dark matter and $\Omega_M
\gg \Omega_B$, dark-matter perturbations
begin growing at matter -- radiation equality
and continue to grow until the present epoch, or nearly so, leading
to significantly more growth and making the observed large-scale
structure consistent with the size of the density perturbations
inferred from CBR anisotropy.

Second, the transition from radiation domination at early
times to matter domination at around 10,000\,yrs leaves its mark on
the shape of the present power spectrum of density
perturbations, and the redshift of matter -- radiation equality
depends upon $\Omega_M$.
Measurements of the shape of the present power spectrum based upon redshift
surveys indicate that the shape parameter $\Gamma = \Omega_M h
\simeq 0.25\pm 0.05$ (see e.g., Peacock \& Dodds, 1994).  For
$h\sim 2/3$, this implies $\Omega_M \sim 0.4$.
(If there are relativistic particles beyond the CBR photons
and relic neutrinos, the formula for the shape parameter changes and
$\Omega_M\gg 0.4$ can be accommodated; see Dodelson et al, 1996).

\subsection{Mass-to-light ratios and $\Omega_M$:
amazingly, the glass is half full!}

The most mature approach to estimating the matter
density involves the use of mass-to-light ratios, the measured
luminosity density, and the deceptively simple equation,
\begin{equation}
\langle \rho_M \rangle = \langle M/L \rangle \, {\cal L}\,,
\end{equation}
where ${\cal L}= 2.4h\times 10^8\,L_{B\odot}\,
{\rm Mpc^{-3}}$ is the measured (B-band) luminosity density of the Universe.
Once the average mass-to-light ratio for
the Universe is determined, $\Omega_M$ follows
by dividing it by the critical mass-to-light ratio,
$(M/L)_{\rm crit} = 1200h$ (in solar units).  Though it is
tantalizingly simple -- and it is far too easy to take any measured
mass-to-light ratio and divide it by $1200h$ -- this method
does not provide an easy and reliable method of determining $\Omega_M$.

Based upon the mass-to-light ratios of the bright, inner regions of
galaxies, $(M/L)_*\sim $\ few, the fraction of critical density in
stars (and closely
related material) has been determined,
$\Omega_* \simeq (0.003\pm 0.001)h^{-1}$ (see e.g., Faber \& Gallagher,
1979).  Persic \& Salucci (1992) derive a similar value
based upon the observed stellar-mass function.
Luminous matter accounts for only a tiny fraction of the
total mass density and only about a tenth of the baryons.

CNOC (Carlberg et al, 1996, 1997) have done a
very careful job of determining a mean cluster mass-to-light
ratio, $(M/L)_{\rm cluster} = 240\pm 50$, which translates
to an estimate of the mean matter density,
$\Omega_{\rm cluster} = 0.20 \pm 0.04$.
Because clusters contain thousands of galaxies and cluster
galaxies do not seem {\em radically} different from field galaxies,
one is tempted to take this estimate of the mean matter density
seriously.  However, it is significantly smaller than the
value I advocated earlier, $\Omega_M = 0.4\pm 0.1$.  Which
estimate is right?

\begin{figure}
\centerline{\psfig{figure=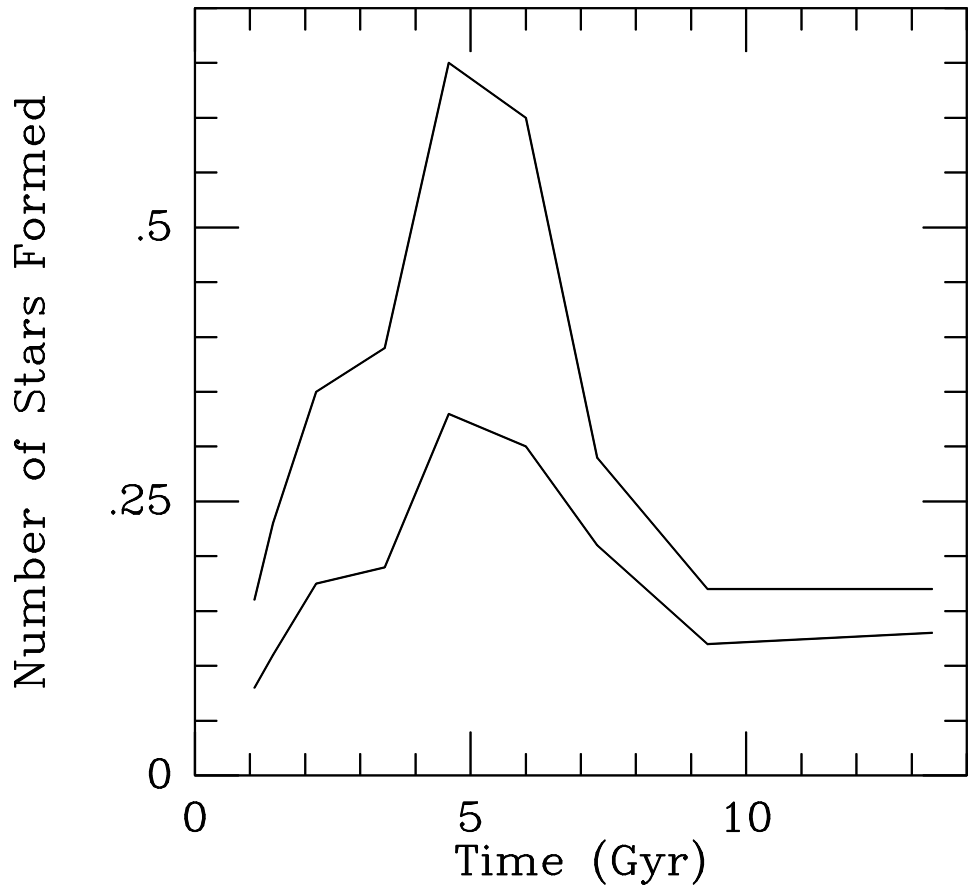,width=5in}}
\caption{Star formation history as a function
of cosmic time (adapted from Madau, 1999).  The
two curves bracket the estimated uncertainty
and the ordinate is given by the star formation
rate times cosmic time.  Note that today we are on
the tail end of star formation.
}
\label{fig:sfr}
\end{figure}

I believe the higher number, based upon the cluster baryon
fraction, is more reliable and that we should be surprised that
the CNOC number is so close, closer than we had any right to expect!
After all, only a small fraction of galaxies
are found in clusters and the luminosity density ${\cal L}$ itself
evolves strongly with redshift and corrections for this effect
are large and uncertain.  (We are on the tail end of star formation
in the Universe:  80\% of star formation took place
at a redshift greater than unity; see Fig.~\ref{fig:sfr}.)
While the value for $\Omega_M$ derived from the cluster baryon
fraction also relies upon clusters, the underlying assumption
is far weaker and much more justified, namely that clusters provide
a fair sample of matter in the Universe.

Even if mass-to-light ratios were measured in the red (they typically are not),
where the light is dominated by low-mass stars and reflects
the integrated history of star formation
rather than the current star-formation rate as it does in the blue,
one would still require the fraction of baryons converted into stars in
clusters to be identical to that in the field to have agreement
between the CNOC estimate and that based upon the cluster baryon
fraction.  Apparently, the fraction of baryons converted into
stars in the field and in clusters is similar, but not identical.

To put this in perspective and to emphasize the shortcomings
of the mass-to-light technique, had one used the cluster mass-to-x-ray ratio
and the x-ray luminosity density, one would have inferred $\Omega_M
\sim 0.05$.  A factor of two discrepancy based upon
optical mass-to-light ratios does not seem so bad.  Enough said.

\subsection{Missing energy?}

The results $\Omega_0 = 1\pm 0.2$ and $\Omega_M = 0.4\pm 0.1$
are in apparent contradiction, suggesting that one or both are
wrong.  However, prompted by a strong
belief in a flat Universe, theorists have explored the
remaining logical possibility:
a dark, exotic form of energy that is smoothly distributed and
contributes 60\% of the critical density
(Turner et al, 1984; Peebles, 1984).  Being
smoothly distributed its presence would not have been
detected in measurements of the matter density.  The properties
of this missing energy are severely constrained by other cosmological
facts, including structure formation, the age of the Universe,
and CBR anisotropy.  So much so, that a smoking-gun signature
for the missing energy was predicted (see e.g., Turner, 1991).

\begin{figure}
\centerline{\psfig{figure=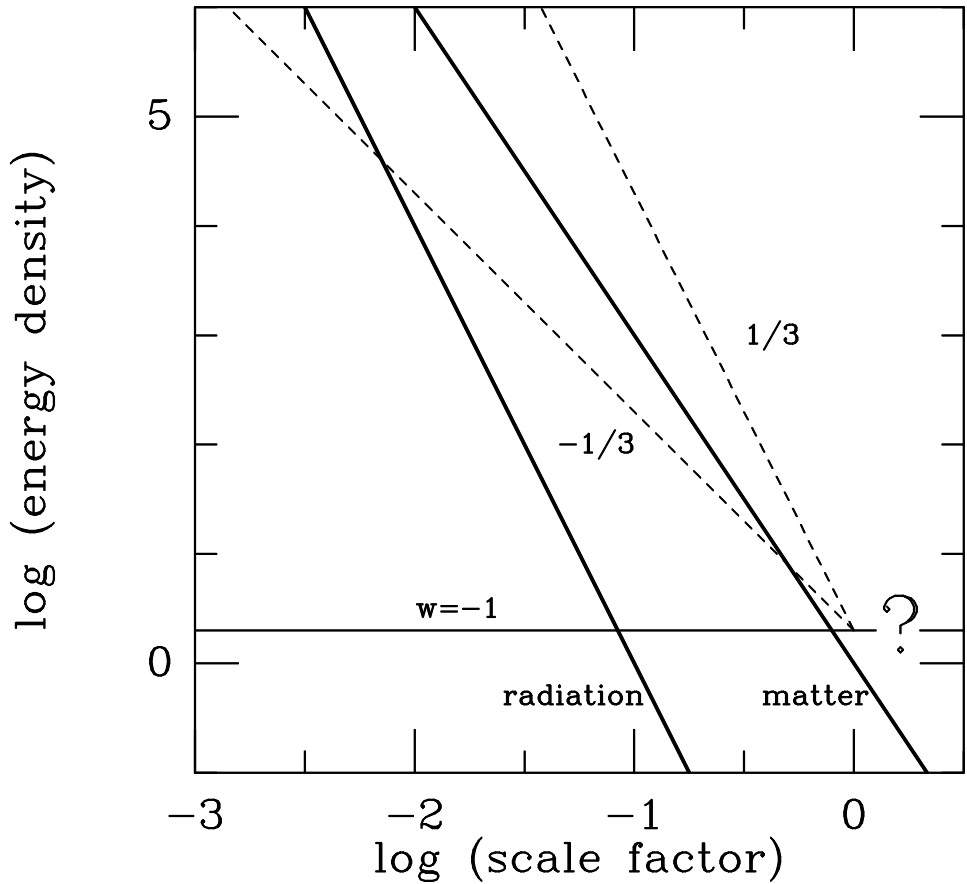,width=5in}}
\caption{Evolution of the energy density in matter,
radiation (heavy lines), and different possibilities
for the dark-energy component ($w=-1,-{1\over 3},{1\over 3}$)
vs. scale factor.  The matter-dominated era begins
when the scale factor was $\sim 10^{-4}$ of its present
size (off the figure) and ends when the dark-energy
component begins to dominate, which depends upon the
value of $w$:  the more negative $w$ is, the longer
the matter-dominated era in which density perturbations
can go into the large-scale structure seen today.
These considerations require $w<-{1\over 3}$ (Turner
\& White, 1997).
}
\label{fig:xmatter}
\end{figure}

To begin, let me parameterize the bulk equation of state of
this unknown component:  $w = p_X/\rho_X$.  This implies
that its energy density evolves as $\rho_X \propto
R^{-n}$ where $n=3(1+w)$.  The development
of the structure observed today from density perturbations of the
size inferred from measurements of the anisotropy of the CBR
requires that the Universe be matter dominated from the epoch
of matter -- radiation equality until very recently.  Thus,
to avoid interfering with structure formation, the dark-energy component
must be less important in the past than it is today.
This implies that $n$ must be less than $3$ or $w< 0$; the more negative
$w$ is, the faster this component gets out of the way (see
Fig.~\ref{fig:xmatter}).  More careful consideration of the
growth of structure implies that $w$ must be less than about
$-{1\over 3}$ (Turner \& White, 1997).

\begin{figure}
\centerline{\psfig{figure=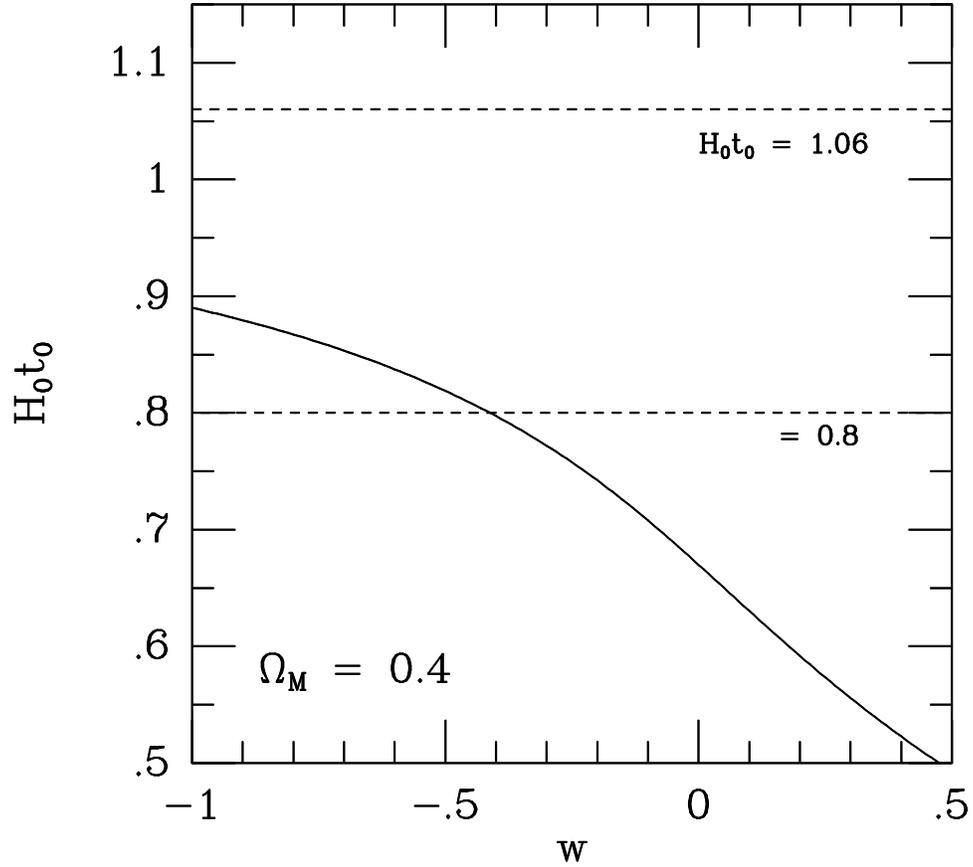,width=5in}}
\caption{$H_0t_0$ vs. the equation of state for the
dark-energy component.  As can be seen, an added benefit of
a component with negative pressure is an older Universe for a given
Hubble constant.  The broken horizontal lines denote the
$1\sigma$ range for $H_0=65\pm 5\,{\rm km\,s^{-1}\,Mpc^{-1}}$
and $t_0=14\pm 1.5\,$Gyr, and indicate that $w<-{1\over 2}$ is
preferred.
}
\label{fig:wage}
\end{figure}

Next, consider the constraint provided by the age of the Universe
and the Hubble constant.  Their product, $H_0t_0$, depends the
equation of state of the Universe; in particular, $H_0t_0$ increases with
decreasing $w$ (see Fig.~\ref{fig:wage}).  To be definite, I will take $t_0
=14\pm 1.5\,$Gyr and $H_0=65\pm 5\,{\rm km\,s^{-1}\,Mpc^{-1}}$
(see e.g., Chaboyer et al, 1998 and Freedman, 1999); this implies
that $H_0t_0 = 0.93 \pm 0.13$.  Fig.~\ref{fig:wage} shows that
$w<-{1\over 2}$ is preferred by age/Hubble constant considerations.

To summarize, consistency between $\Omega_M \sim 0.4$ and
$\Omega_0 \sim 1$ along with other cosmological considerations
implies the existence of a dark-energy component with bulk
pressure more negative than about $-\rho_X /2$.  The simplest
example of such is vacuum energy (Einstein's cosmological
constant), for which $w=-1$.  The smoking-gun signature of
a smooth, dark-energy component is accelerated expansion since
$q_0 = 0.5 + 1.5w\Omega_X \simeq 0.5 + 0.9w < 0$ for
$w<-{5\over 9}$.

\subsection{Missing energy found!}

\begin{figure}
\centerline{\psfig{figure=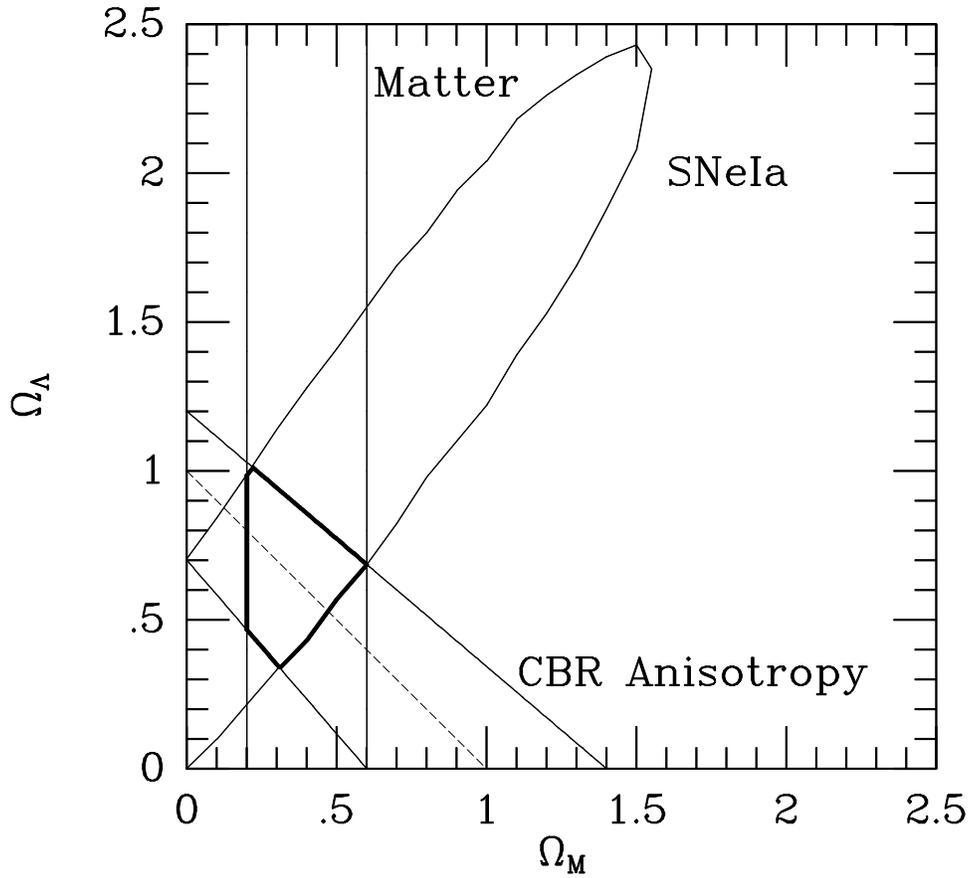,width=5in}}
\caption{Two-$\sigma$ constraints to $\Omega_M$ and $\Omega_\Lambda$
from CBR anisotropy, SNe Ia, and measurements of clustered matter.
Lines of constant $\Omega_0$ are diagonal, with a flat
Universe shown by the broken line.
The concordance region is shown in bold:  $\Omega_M\sim 1/3$,
$\Omega_\Lambda \sim 2/3$, and $\Omega_0 \sim 1$.
(Particle physicists who rotate the figure by $90^\circ$
will recognize the similarity to the convergence of the
gauge coupling constants.)
}
\label{fig:omegalambda}
\end{figure}

In 1998 evidence for accelerated expansion was presented in the form
of the magnitude -- redshift (Hubble)
diagram for fifty-some type Ia supernovae (SNe Ia)
out to redshifts of nearly 1.
Two groups, the Supernova Cosmology Project (Perlmutter et al, 1998;
Goobar, 1999) and the High-z Supernova Search Team (Riess et al, 1998;
Schmidt et al, 1998), working independently and using different
methods of analysis, each found evidence for accelerated expansion.
Perlmutter et al (1998) summarize their results as a constraint
to a cosmological constant (see Fig.~\ref{fig:omegalambda}),
\begin{equation}
\Omega_\Lambda = {4\over 3}\Omega_M +{1\over 3} \pm {1\over 6}\,.
\end{equation}
For $\Omega_M\sim 0.4 \pm 0.1$, this implies $\Omega_\Lambda =
0.85 \pm 0.2$, or just what is needed to account for the missing energy!

(A simple explanation of the SN Ia results may be
useful.  If galactic distances and velocities were measured today they
would obey a perfect Hubble law, $v_0=H_0d$, because the expansion
of the Universe is simply a rescaling.  Because we see distant galaxies
at an earlier time, their velocities should be larger than predicted
by the Hubble law, provided the expansion is slowing due to the
attractive force of gravity.
Using SNe Ia as standard candles to determine the distances
to faraway galaxies, the two groups in effect found the opposite:
distant galaxies are moving slower than predicted by the Hubble
law, implying the expansion is speeding up!)
                                            
Recently, two other studies, one based upon the x-ray properties of
rich clusters of galaxies (Mohr et al, 1999) and the other based
upon the properties of double-lobe radio galaxies
(Guerra et al, 1998), have reported evidence
for a cosmological constant (or similar dark-energy component)
that is consistent with the SN Ia results (i.e., $\Omega_\Lambda \sim 0.7$).

There is another powerful test of an accelerating Universe whose
results are more ambiguous.  It is based upon the fact that
the frequency of multiply lensed
QSOs is expected to be significantly higher in an accelerating
universe (Turner, 1990).  Kochanek (1996) has used gravitational
lensing of QSOs to place a
95\% cl upper limit, $\Omega_\Lambda < 0.66$; and Waga and Miceli
(1998) have generalized it to a dark-energy component with negative
pressure:  $\Omega_X < 1.3 + 0.55w$ (95\% cl), both results for a flat
Universe.  On the other hand, Chiba and Yoshii (1998) claim evidence
for a cosmological constant, $\Omega_\Lambda = 0.7^{+0.1}_{-0.2}$,
based upon the same data.  From this I conclude:  1) lensing
excludes $\Omega_\Lambda$ larger than 0.8, and 2) when
larger objective surveys of gravitational-lensed
QSOs are carried out (e.g., the Sloan Digital Sky Survey),
there is the possibility of uncovering another smoking-gun
for accelerated expansion.

By far, the strongest evidence for dark energy is the SN Ia data.
The statistical errors reported by the two groups are smaller
than possible systematic errors.  Thus, the believability of
the results turn on the reliability
of SNe Ia as one-parameter standard candles.  SNe Ia are thought
to be associated with the nuclear detonation of Chandrasekhar-mass
white dwarfs.  The one parameter is the rate of decline
of the light curve:  The brighter ones decline more slowly (the
so-called Phillips relation; see Phillips, 1993).  The
lack of a good theoretical understanding of this
(e.g., what is the physical parameter?) is offset
by strong empirical evidence for the relationship
between peak brightness and rate of decline, based upon a sample
of nearby SNe Ia.  It is reassuring that in all respects
studied, the distant sample of SNe Ia appear to be similar to
the nearby sample.  For example, distribution of decline rates
and dispersion about the Phillips relationship.  The local
sample spans a range of metallicity, likely
spanning that of the distant sample, and further, suggesting that
metallicity is not an important second parameter.

At this point, it is fair to say that if there is a problem with SNe Ia
as standard candles, it must be subtle.  Cosmologists are even more
inclined to believe the SN Ia results
because of the preexisting evidence for a ``missing-energy component''
that led to the prediction of accelerated expansion.

\subsection{Cosmic concordance}

\begin{figure}
\centerline{\psfig{figure=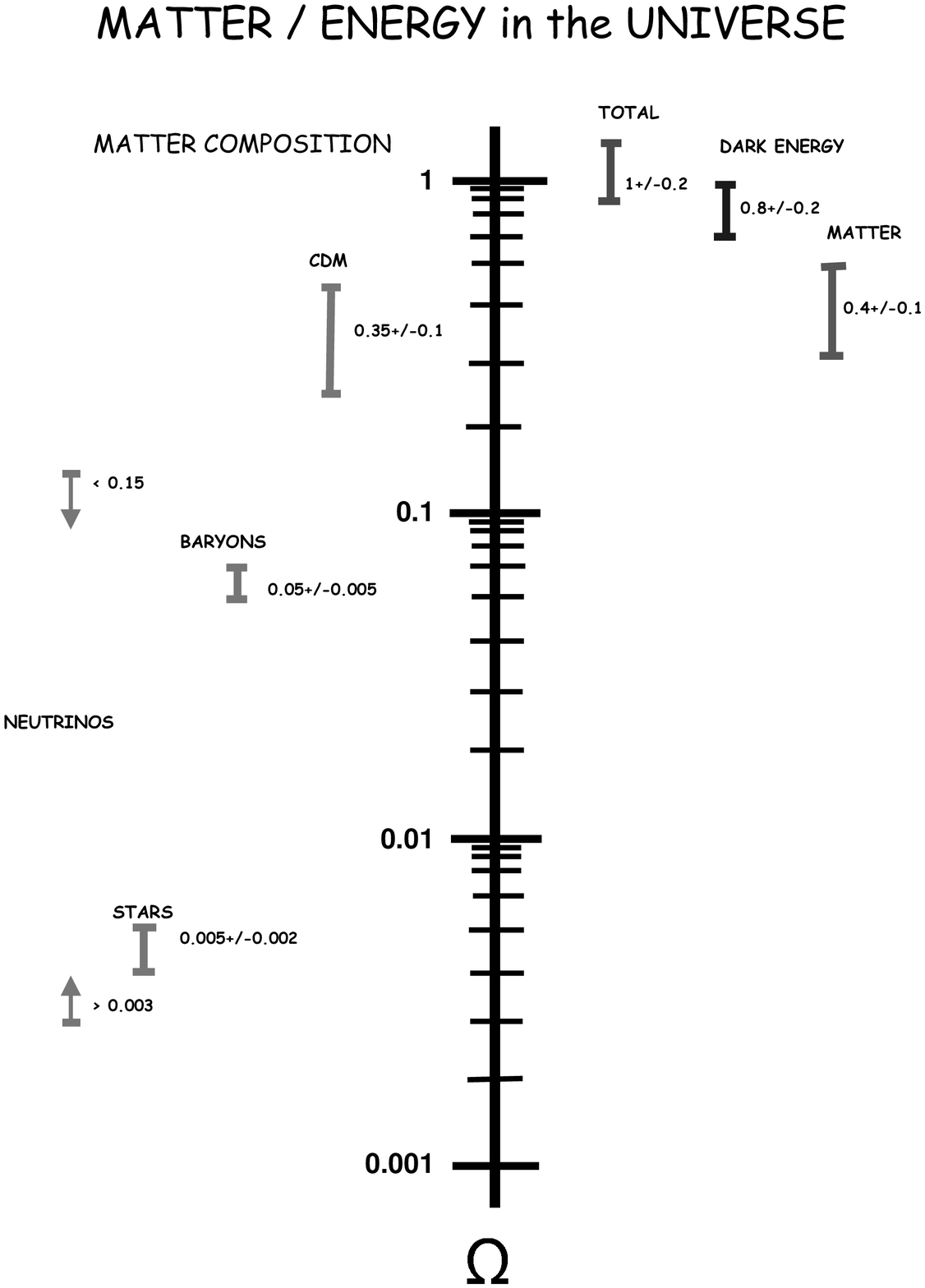,width=5in}}
\caption{Summary of matter/energy in the Universe.
The right side refers to an overall accounting of matter
and energy; the left refers to the composition of the matter
component.  The contribution of relativistic particles,
CBR photons and neutrinos, $\Omega_{\rm rel}h^2 = 4.170\times
10^{-5}$, is not shown.  The upper limit to mass density contributed
by neutrinos is based upon the failure of the hot dark
matter model of structure formation (White, Frenk \& Davis,
1983; and Dodelson et al, 1996) and the lower limit follows from the
evidence for neutrino oscillations (Fukuda et al, 1998).
Here $H_0$ is taken to be $65\,{\rm km\,s^{-1}\,Mpc^{-1}}$.
}
\label{fig:omega}
\end{figure}

With the SN Ia results we have for the first time a complete
and self-consistent
accounting of mass and energy in the Universe (see Fig.~\ref{fig:omega}).
The consistency of the matter/energy accounting
is illustrated in Fig.~\ref{fig:omegalambda}.
Let me explain this exciting figure.  The SN Ia results are sensitive to the
acceleration (or deceleration) of the expansion
and constrain the combination ${4\over 3}\Omega_M -\Omega_\Lambda$.  (Note,
$q_0 = {1\over 2}\Omega_M - \Omega_\Lambda$; ${4\over 3}\Omega_M -
\Omega_\Lambda$ corresponds to the deceleration parameter
at redshift $z\sim 0.4$, the median redshift of these
samples).  The (approximately) orthogonal combination,
$\Omega_0 = \Omega_M + \Omega_\Lambda$
is constrained by CBR anisotropy.  Together, they define a concordance
region around $\Omega_0\sim 1$, $\Omega_M \sim 1/3$,
and $\Omega_\Lambda \sim 2/3$.  The constraint
to the matter density alone, $\Omega_M = 0.4\pm 0.1$,
provides a cross check, and it is consistent with these numbers.
Cosmic concordance!

\begin{figure}
\centerline{\psfig{figure=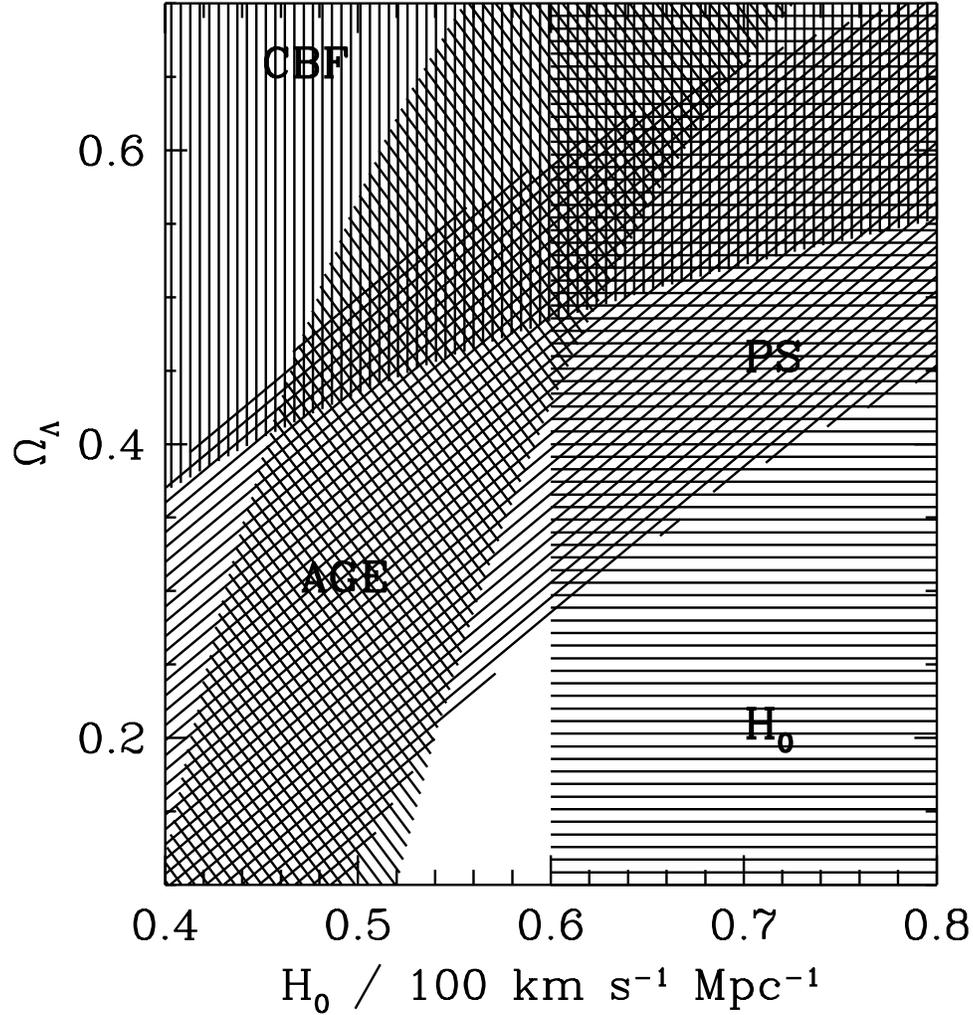,width=5in}}
\caption{Constraints used to determine the best-fit CDM model:
PS = large-scale structure + CBR anisotropy; AGE = age of the
Universe; CBF = cluster-baryon fraction; and $H_0$= Hubble
constant measurements.  The best-fit model, indicated by
the darkest region, has $H_0\simeq 60-65\,{\rm km\,s^{-1}
\,Mpc^{-1}}$ and $\Omega_\Lambda
\simeq 0.55 - 0.65$.  Evidence for its smoking-gun signature --
accelerated expansion -- was presented in 1998 (adapted
from Krauss \& Turner, 1995 and Turner, 1997).}
\label{fig:best_fit}
\end{figure}

But there is more.  We also have a consistent and well motivated
picture for the formation of structure in the Universe, $\Lambda$CDM.
The $\Lambda$CDM model, which is the cold dark
matter model with $\Omega_B \sim 0.05$, $\Omega_{\rm CDM}\sim
0.35$ and $\Omega_\Lambda \sim 0.6$, is a very good fit to all
cosmological constraints:  large-scale structure, CBR anisotropy,
age of the Universe, Hubble constant and the constraints
to the matter density and cosmological constant; see Fig.~\ref{fig:best_fit}
(Krauss \& Turner, 1995; Ostriker \& Steinhardt, 1995;
Turner, 1997).  Further, as
can be seen in Figs.~\ref{fig:cbr_today} and \ref{fig:cbr_knox},
CBR anisotropy measurements are beginning to show evidence for
the acoustic peaks characteristic of the Gaussian, curvature
perturbations predicted by inflation.
Until 1998, $\Lambda$CDM's only major flaw was the absence
of evidence for accelerated expansion.  Not now.

\section{Three Dark Matter Problems}

While stars are very interesting and pretty to look at --
and without them, astronomy wouldn't be astronomy and we
wouldn't exist -- they
represent a tiny fraction of the cosmic mass budget, only about 0.5\%
of the critical density.  As we have known for several decades,
the bulk of the matter and energy in the Universe
is dark.  The present accounting defines three dark matter/energy
problems; none is yet fully addressed.

\subsection{Dark Baryons}

By a ten to one margin, the bulk of the baryons are dark and not in
the form of stars.  With the exception of clusters, where the
``dark'' baryons exist as hot, x-ray emitting
intracluster gas, the nature of the dark baryons is not known.
Clusters only account for around 10\% or so of the
baryons in the Universe (Persic \& Salucci, 1992)
and the (optically) dark baryons elsewhere,
which account for 90\% or more of all the baryons,
could take on a different form.

The two most promising possibilities for the dark baryons are
diffuse hot gas and ``dark stars''
(white dwarfs, neutron stars, black holes or objects of mass around
or below the hydrogen-burning limit).  I favor the former
possibility for a number of reasons.  First, that's
where the dark baryons in clusters are.  Second, the cluster baryon
fraction argument can be turned around to infer $\Omega_{\rm gas}$
at the time clusters formed, redshifts $z\sim 0 -1$,
\begin{equation}
\Omega_{\rm gas}h^2 = f_{\rm gas} \Omega_Mh^2 =
0.023\,(\Omega_M/0.4)(h/0.65)^{1/2}\,.
\end{equation}
That is, at the time clusters formed, the mean gas density was essentially
equal to the baryon density (unless $\Omega_Mh^{1/2}$ is very small),
thereby accounting for the bulk of baryons in gaseous form.
Third, numerical simulations suggest that most of
the baryons should still be in gaseous form today (Rauch et al, 1997;
Ostriker, 1999).

There are two arguments for dark stars as the baryonic dark matter.
First, the gaseous baryons not associated with clusters have not been
detected.  Second, the results of the microlensing surveys toward the
LMC and SMC (Spiro, 1999) are consistent with about one-third of our halo
being in the form of half-solar mass white dwarfs.

I find neither argument compelling;
gas outside clusters will be cooler ($T\sim 10^5 - 10^6$\,K)
and difficult to detect, either in absorption or emission.
There are equally attractive explanations for the Magellanic
Cloud microlensing events (e.g., self lensing by the Magellanic
Clouds, lensing by stars in the spheroid, or lensing due to
disk material that, due to flaring and warping of the disk,
falls along the line of sight to the LMC; see Sahu, 1994;
Evans et al, 1998; Gates et al, 1998; Zaritsky \& Lin, 1997;
Zhao, 1998).
The white-dwarf interpretation for the halo has a host of troubles:
Why haven't the white dwarfs been seen (Graff et al, 1998)?  The
star formation rate required to produce these white dwarfs --
close to $100\,{\rm yr^{-1}\,Mpc^{-3} }$ -- far exceeds
that measured at any time in the past or present (see Madau, 1999).
Where are the lower-main-sequence stars associated with this stellar
population and the gas, expected to be 6 to 10 times that
of the white dwarfs, that didn't form into stars (Fields et al,
1997)?  Finally, there is evidence that the
lenses for both SMC events are stars within the SMC (Alcock et al,
1998; EROS Collaboration, 1998a,b)
and at least one of the LMC events is explained by an LMC lens.

The SMC/LMC microlensing puzzle can be stated another way.
The lenses have all the characteristics of ordinary,
low-mass stars (e.g., mass and binary frequency).  If this is so,
they cannot be in the halo (they would have been seen);
the puzzle is to figure out where they are located.

\subsection{Cold Dark Matter}

The second dark-matter problem follows from the inequality
$\Omega_M\simeq 0.4 \gg \Omega_B\simeq 0.05$:
There is much more matter than
there are baryons, and thus, nonbaryonic dark matter is
the dominant form of matter.  The evidence for this very
profound conclusion has been mounting for almost two decades, and
this past year, the Burles -- Tytler deuterium measurement anchored the
baryon density and allowed the cleanest determination of
the matter density, through the cluster baryon fraction.

Particle physics provides an attractive solution to the
nonbaryonic dark matter problem:  relic elementary particles
left over from the big bang (see Ellis, 1999).
Long-lived or stable particles
with very weak interactions can remain from the earliest
moments of particle democracy in sufficient numbers to account
for a significant fraction of critical density (very weak interactions
are needed so that their annihilations cease before their
numbers are too small).  The three most promising candidates
are a neutrino of mass 30\,eV or so (or $\sum_i\,m_{\nu_i}
\sim 30\,$eV), an axion of mass
$10^{-5\pm 1}\,$eV, and a neutralino of mass between $50\,$GeV and $500\,$GeV.
All three are motivated by particle-physics theories that attempt
to unify the forces and particles of Nature.  The fact that
such particles can also account for the nonbaryonic dark matter
is either a big coincidence or a big hint.  Further, the fact
that these particles interact with each other and ordinary matter
very weakly, provides a simple and natural explanation for dark matter
being more diffusely distributed.

At the moment, there is significant circumstantial
evidence against neutrinos as the bulk of the dark matter.  Because
they behave as hot dark matter, structure forms from
the top down, with superclusters fragmenting into clusters
and galaxies (White, Frenk \& Davis, 1983),
in stark contrast to the observational
evidence that indicates structure formed from the bottom
up.  (Hot + cold dark matter is still an outside possibility,
with $\Omega_\nu\sim 0.15$ or less; see Dodelson et al, 1996
and Gawiser \& Silk, 1998.)
Second, the evidence for neutrino mass based upon
the atmospheric (Totsuka, 1999) and solar-neutrino (Kirsten, 1999
and J. Bahcall, 1999) data suggests a
neutrino mass pattern with the tau neutrino at $0.1\,$eV,
the muon neutrino at $0.001\,$eV to $0.01$\,eV and the
electron neutrino with an even smaller mass.  In particular,
the factor-of-two deficit of atmospheric muons neutrinos with
its dependence upon zenith angle is very strong evidence
for a neutrino mass difference
squared between two of the neutrinos of around $10^{-2}$\,eV$^2$
(Fukuda et al, 1998).  This sets a {\em lower} bound to
neutrino mass of about $0.1\,$eV, implying neutrinos contribute
at least as much mass as bright stars.  WOW!

Both the axion and neutralino behave as cold dark matter; the
success of the cold dark matter model of structure formation
makes them the leading particle dark-matter candidates.  Because
they behave as cold dark matter, they are expected to be the
dark matter in our own halo; in fact, there is nothing that
can keep them out (Gates \& Turner, 1994).  As discussed above,
2/3 of the dark matter in our halo -- and probably all the
halo dark matter -- cannot be explained by baryons in any form.
The local density of halo material is estimated to be $10^{-24}\,{\rm g\,
cm^{-3}}$, with an uncertainty of slightly less than a factor
of 2 (Gates et al, 1995).  This makes the halo of our galaxy
an ideal place to look for cold dark matter particles!
An experiment at Livermore National Laboratory with sufficient sensitivity
to detect halo axions is currently taking data (van Bibber et al,
1998; Rosenberg and van Bibber, 1999) and experiments
at several laboratories around the
world are beginning to search for halo neutralinos with sufficient sensitivity
to detect them (Sadoulet, 1999).  The particle dark-matter hypothesis
is compelling, but very bold, and most importantly,
it is now being tested.

Finally, while the axion and the neutralino are the most promising
particle dark-matter candidates, neither one is a ``sure thing.''
Moreover, any sufficiently heavy particle relic (mass greater than a GeV
or so) will behave as cold dark matter.  A host of more exotic
possibilities have been suggested, from solar-mass primordial black holes
produced at the quark/hadron transition (see e.g., Jedamzik, 1998
and Jedamzik \& Niemeyer, 1998) that masquerade as MACHOs in our
halo to supermassive (mass greater than $10^{10}\,$GeV) particles
produced by nonthermal processes at the end of inflation (see e.g.,
Kolb, 1999).  Lest we become overconfident, we should remember
that Nature has many options for the particle dark matter.

\subsection{Dark Energy}

I have often used the term exotic to refer to particle
dark matter.  That term will now have to be reserved for the
dark energy that is causing the accelerated expansion of the
Universe -- by any standard, it is more exotic and more
poorly understood.  Here is what we do know:   it contributes
about 60\% of the critical density; it has pressure more negative
than about $-\rho /2$; and it does not clump
(otherwise it would have contributed to estimates
of the mass density).  The simplest possibility is
the energy associated with the virtual particles that populate
the quantum vacuum; in this case $p=-\rho$ and the dark energy
is absolutely spatially and temporally uniform.

This ``simple'' interpretation has its difficulties.   Einstein ``invented''
the cosmological constant to make a static model of the Universe
and then he discarded it; we now know that the concept is not optional.
The cosmological constant corresponds to the energy associated
with the vacuum.  However, there is no sensible calculation of
that energy (see e.g., Zel'dovich, 1967; Bludman and Ruderman,
1977; and Weinberg, 1989),
with estimates ranging from $10^{122}$ to $10^{55}$ times the critical
density.  Some particle physicists believe that when the
problem is understood, the answer will be zero.  Spurred
in part by the possibility that cosmologists may have actually weighed
the vacuum (!), particle theorists are taking a fresh look
at the problem (see e.g., Harvey, 1998; Sundrum, 1997).  Sundrum's
proposal, that the energy of the vacuum is close to the present
critical density because the graviton is a composite particle
with size of order 1\,cm, is indicative of the profound
consequences that a cosmological constant has for fundamental physics.

Because of the theoretical problems mentioned above, as well as the
checkered history of the cosmological constant, theorists have
explored other possibilities for a smooth, component to the dark
energy (see e.g., Turner \& White, 1997).
Wilczek and I pointed out that even if the
energy of the true vacuum is zero, as the Universe as
cooled and went through a series of phase transitions, it
could have become hung up in a metastable vacuum with
nonzero vacuum energy (Turner \& Wilczek, 1982).  In the
context of string theory, where there are a very large number
of energy-equivalent vacua, this becomes a more
interesting possibility:  perhaps the degeneracy of vacuum
states is broken by very small effects, so small that
we were not steered into the lowest energy vacuum during
the earliest moments.

Vilenkin (1984) has suggested a tangled network of very light
cosmic strings (also see, Spergel \& Pen, 1997) produced
at the electroweak phase transition; networks of other frustrated
defects (e.g., walls) are also possible.  In general, the
bulk equation-of-state of frustrated defects
is characterized by $w=-N/3$ where $N$
is the dimension of the defect ($N=1$ for strings, $=2$
for walls, etc.).  The SN Ia data almost exclude strings,
but still allow walls.

An alternative that has received a lot of attention is
the idea of a ``decaying cosmological constant'', a termed
coined by the Soviet cosmologist Matvei Petrovich Bronstein in 1933
(Bronstein, 1933).  (Bronstein was executed on Stalin's
orders in 1938, presumably for reasons not directly related to the
cosmological constant; see Kragh, 1996.)
The term is, of course, an oxymoron; what people have in mind
is making vacuum energy dynamical.  The simplest realization
is a dynamical, evolving scalar field.  If it is spatially homogeneous,
then its energy density and pressure are given by
\begin{eqnarray}
\rho & = & {1\over 2}{\dot\phi}^2 + V(\phi ) \nonumber \\
p    & = & {1\over 2}{\dot\phi}^2 - V(\phi )
\end{eqnarray}
and its equation of motion by (see e.g., Turner, 1983)
\begin{equation}
\ddot \phi + 3H\dot\phi + V^\prime (\phi ) = 0
\end{equation}

The basic idea is that energy of the true vacuum is zero, but
not all fields have evolved to their state of minimum
energy.  This is qualitatively different from that
of a metastable vacuum, which is a local minimum of the
potential and is classically stable.  Here, the field is
classically unstable and is rolling toward its lowest
energy state.

Two features of the ``rolling-scalar-field scenario'' are
worth noting.  First, the effective equation of state,
$w=({1\over 2}\dot\phi^2 - V)/({1\over 2}\dot\phi^2 +V)$,
can take on any value from 1 to $-1$.  Second, $w$ can
vary with time.  These are key features that may allow it
to be distinguished from the other possibilities.  In fact,
there is some hope that more SNe Ia will be able to do
this and perhaps even permit the reconstruction of the scalar-field
potential (Huterer \& Turner, 1998).

The rolling scalar field scenario (aka mini-inflation
or quintessence) has received a lot of attention over
the past decade (Freese et al, 1987; Ozer \& Taha, 1987;
Ratra \& Peebles, 1988; Frieman et al, 1995; Coble et al, 1996;
Turner \& White, 1997; Caldwell et al, 1998; Steinhardt, 1999).
It is an interesting idea,
but not without its own difficulties.  First,
one must {\em assume} that the energy of the
true vacuum state ($\phi$ at the minimum of its potential)
is zero; i.e., it does not address the cosmological
constant problem.  Second, as Carroll (1998) has emphasized,
the scalar field is very light and can mediate long-range forces.
This places severe constraints on it.  Finally,
with the possible exception of one model (Frieman et al,
1995), none of the
scalar-field models address how $\phi$ fits into the
grander scheme of things and why it is so light ($m\sim 10^{-33}\,$eV).

\section{Concluding Remarks}

1998 was a very good year for cosmology.  We now have
a plausible and complete accounting
of matter and energy in the Universe; in $\Lambda$CDM,
a model for structure formation that is consistent with
all the data at hand; and the first evidence for the key
tenets of inflation (flat Universe and adiabatic
density perturbations).  One normally conservative cosmologist
has gone out on a limb by stating that 1998 may be a turning point in
cosmology as important as 1964, when the CBR was discovered (Turner, 1999).

We still have important questions to address:  Where
are the dark baryons?  What is the dark matter?  What
is the nature of the dark energy?  What is the explanation for the
complicated pattern of mass and energy:
neutrinos (0.3\%), baryons (5\%),
cold dark matter particles (35\%) and dark energy (60\%)?
Especially puzzling is the ratio of dark energy to dark matter:
because they evolve differently with time, the ratio of dark matter
to dark energy was higher in the past and will be smaller in
the future; only today are they comparable.  WHY NOW?

While we have many urgent questions, we can see
a flood of precision cosmological and laboratory
data coming that will help to answer these questions:
High-resolution maps of CBR anisotropy (MAP and Planck);
large redshift surveys (SDSS and 2dF); more SN Ia data;
experiments to directly detect halo axions and neutralinos;
more microlensing data (MACHO, EROSII, OGLE, AGAPE, and
superMACHO); accelerator experiments at Fermilab and CERN,
searching for the neutralino and its supersymmetric friends
and further evidence for neutrino mass, and at the KEK
and SLAC B-factories, revealing more about the nature
of $CP$ violation; and nonaccelerator experiments that will
shed further light on neutrino mass, particle dark matter,
new forces, and the nature of gravity.

These are exciting times in cosmology!

\paragraph{Acknowledgments.}
This work was supported by the DoE (at Chicago and Fermilab) and by
the NASA (at Fermilab by grant NAG 5-7092).  I wish to thank the organizers,
L. Bergstrom, P. Carlson, and C. Fransson, both for their graciousness
as hosts and for putting on such a fine meeting.

\begin{figure}
\centerline{\psfig{figure=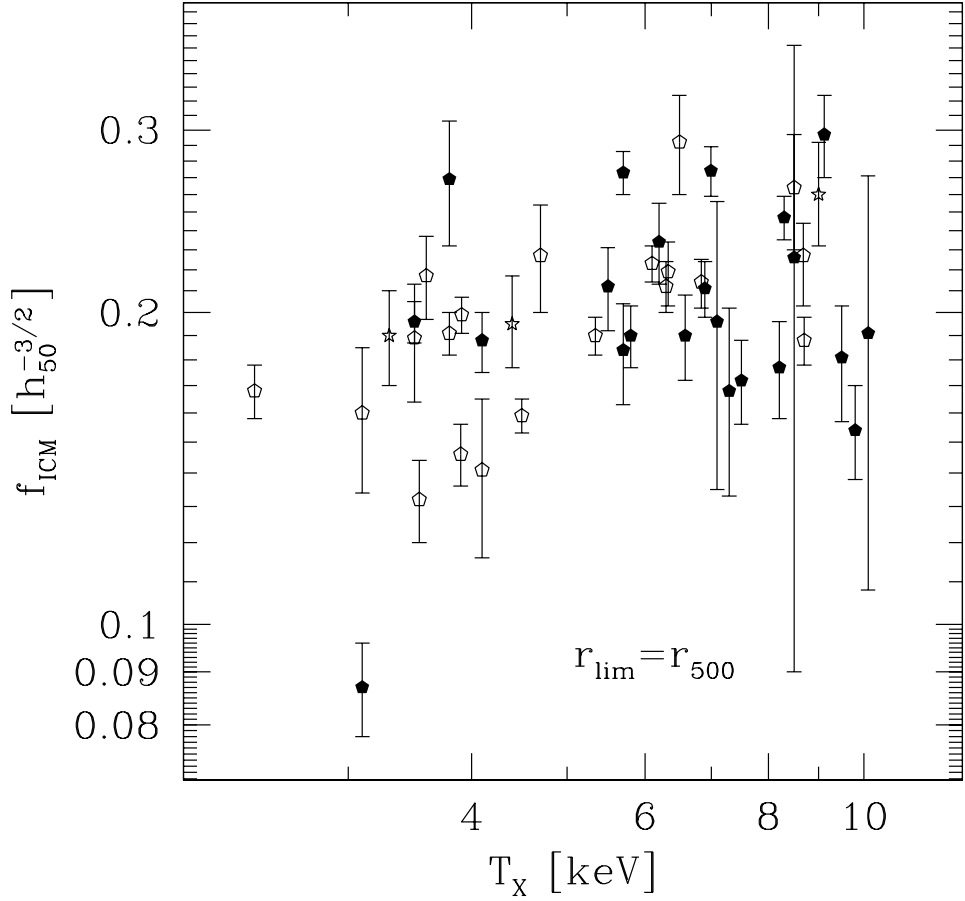,width=5in}}
\caption{Cluster gas fraction as a function of
cluster gas temperature for a sample of 45 galaxy clusters
(Mohr et al, 1998).  While there is some indication
that the gas fraction
decreases with temperature for $T< 5\,$keV, perhaps because
these lower-mass clusters lose some of their hot gas, the
data indicate that the gas fraction reaches a plateau
at high temperatures, $f_{\rm gas} =0.212 \pm 0.006$
for $h=0.5$ (Figure courtesy of Joe Mohr).
}
\label{fig:gas}
\end{figure}

\end{document}